\documentclass[twocolumn]{article}
\usepackage{cite}
\usepackage{amsmath,amssymb,amsfonts}
\usepackage{algorithmic}
\usepackage{graphicx}
\usepackage{textcomp}
\usepackage{xcolor}
\usepackage{tikz}
\usetikzlibrary{shapes.geometric, arrows}

\title{Evaluating Credit VIX (CDS IV) Prediction Methods with Incremental Batch Learning [Preprint]}

\author{
    Robert Taylor \\
    EECS, Queen Mary, University of London \\
    London, UK \\
    \texttt{robert.taylor.94@outlook.com}
}

\date{} % Optional: You can include a date or leave it empty

\begin{document}

\maketitle

\begin{abstract}
This paper presents the experimental process and results of SVM, Gradient Boosting, and an Attention-GRU Hybrid model in predicting the Implied Volatility of rolled-over five-year spread contracts of credit default swaps (CDS) on European corporate debt during the quarter following mid-May '24, as represented by the iTraxx/Cboe Europe Main 1-Month Volatility Index (BP Volatility). The analysis employs a feature matrix inspired by Merton's determinants of default probability. Our comparative assessment aims to identify strengths in \textit{SOTA} and classical machine learning methods for financial risk prediction.
\end{abstract}

\section{Introduction}
The forecasting task being presented is of a single target but is multi-task in application, as it applies to the accurate estimation of aggregated Implied Volatility together with a credit risk component (if we are to accept IV as a risk-neutral market expectation of risk)\cite{Gunnarsson2024}.

We approach this task via the lens of the former, which has broad applications, spanning portfolio optimisation\cite{Mesquita2023}, derivative pricing, risk management\cite{Poon2003}, and econometric modeling more generally. Reliable volatility estimates underpin mean-variance optimization (MVO) \cite{Markowitz1952} and dynamic asset allocation for position sizing and targeting strategies. For derivatives pricing, accurate forecasts of IV are essential for constructing volatility surfaces and estimating fair values using models such as Black-Scholes-Merton (BSM) \cite{Black1973} and Heston \cite{Heston1993} to manage Greeks.

In financial risk management, volatility estimates are crucial for Value at Risk (VaR) \cite{Jorion2006} and Expected Shortfall (ES) calculations, quantifying potential losses within specified confidence intervals. Such models can also be employed for maximum likelihood estimations (MLEs) or in conjunction with bootstrapping to form empirical probability density functions (PDFs). Volatility modeling also supports stress testing and scenario analysis, assessing portfolio resilience under extreme conditions.

High-frequency trading also extensively relies on volatility estimates for real-time strategy adjustments, execution risk management, and arbitrage \cite{Aldridge2013}. Additionally, IV is included in financial stability assessments beyond capital markets, such as macroprudential policymaking \cite{IMF2014}.

As outlined, we take a slight departure from typical IV estimation by conducting forecasts of the IV of the European iTraxx (marketed as the `Credit Vix' by S\&P) \cite{SPGlobal2024}, where the aggregated CDS spreads, which comprise the underlying index, can be considered a measure of systemic risk for the European credit market \cite{Zopounidis2015,Mesquita2023}. The underlying `Europe Main' Index is representative of the 125 most most capitalised companies in the European market, so resulting IV will differ from options on single-name CDS contracts.

CDS are a contingent claim resembling an insurance contract between two market participants wishing to take a view on credit quality, where protection is obtained against the default of an underlying reference entity\cite{Hull2000}. An obvious difficulty in finding features that predict corporate market solvency is that CDS spreads themselves are among the best predictors of corporate market solvency\cite{Blanco2005}.

\subsection{Objectives}
We will of course pose this problem as an Implied Volatility, rather than credit risk, prediction task, given the target's derivation and obvious statistical likeness (high positive skew, endogeneity, clustering, mean-reverting properties and fractal-like path\cite{Mandelbrot1963, Gatheral2018}.

Classical econometric and stochastic volatility (SV) models, such as GARCH \cite{Bollerslev1986} and HAR \cite{Corsi2009}, attempt to handle time-varying volatility and volatility clustering. However, they are limited in their ability to incorporate nonlinear relationships across significant observations temporally or other latent states that machine learning (ML) models are purported to effortlessly extract (conditional on feature extraction) from the rich interplay of covariates. The task, further outlined below, is to contribute to the literature that has challenged such models.

Supervised learning approaches, which tend to make fewer assumptions about the data-generating process (DGP)\cite{Dixon2020}, are increasingly being posited as alternatives to traditional Stochastic Volatility (SV) or classical autoregressive (AR) models in contemporary literature, within the broader exploration of data-driven models as substitutes for closed-form solutions \cite{Chevillon2004}. This study investigates these approaches using market-derived features typically associated with credit risk\cite{Merton1974}, which can serve as substitutes in the absence of order-book level data that is typically required for SV models.

Implied Volatility (IV), unlike Realized Volatility (RV) \cite{Andersen2009} or Historical Volatility (HV), is an \textit{implied} latent variable commonly derived by solving for the volatility parameter, $\sigma$, in the Black-Scholes-Merton (BSM) model or similar. IV represents the volatility value that, when input into such a model, produces an option price congruent with the observed market price. This implies that it also captures discrepancies between the market's expectations and the theoretical assumptions of the model.\cite{Gunnarsson2024}. while indices like the iTraxx/Cboe Europe Main 1-Month Volatility Index do not strictly use BSM, they are similarly derived from option market data using a closed-form model.

The iTraxx/Cboe Europe Main 1-Month Volatility Index (Credit VIX) calculates implied volatility by averaging the variances implied by options on CDS indices. The variances are calculated using a formula of the approximate form:
\vspace{0.01cm}
\small
\[
\sigma^2 \approx \frac{2}{T \times \text{RPV01}} \sum \left(\frac{P(K) \times \Delta K}{K^2}\right) - \frac{1}{T} \left(\frac{\text{CDSI}}{K_0} - 1\right)^2
\]
\normalsize
\vspace{0.5cm}
where the sum represents the aggregation of option prices (\( P(K) \)) over different strike prices (\( K \)) and maturities, weighted by the intervals between strike prices (\( \Delta K \)). Here, RPV01 (Risky Present Value of 1BP) represents the present value of a 1 basis point change in the credit spread of the underlying index, adjusted for the probability of default, and CDSI (CDS Index Spread) is the forward spread of the index, reflecting the market-implied cost of protection against defaults.

This measure of credit spread IV is fully outlined in the Credit VIX Indices Methodology \cite{CreditVix2023}.

\subsection{Contributions}
Our main contributions are:
\begin{itemize}
    \item The application of a GRU Network with Multi-Headed Attention and residual connections in the estimation of risk-neutral credit risk/IV.
    \item Comprehensive blind evaluation on contemporary high-vol regime.
    \item Feature set with economic determinants, which is limited in the literature\cite{Bai2023}.
    \item Detailed assessment of comparative performance over multiple training windows
\end{itemize}

\section{Literature Review}
\label{sec:literature}

\subsection{Existing Approaches in ML}
In rough order of complexity, we examine research surrounding candidate models for this task. We will first visit LightGBM, a panel data prediction technique developed by Microsoft, but an extension of the gradient boosting class of prediction models \cite{Ke2017}. Gradient Boosting has significant support in the time-series prediction space, particularly as a winner of multiple Kaggle competitions \cite{Chen2016}, together with its success in the Monash M5 `Accuracy' Time-Series Forecasting Competitions \cite{Makridakis2022}. Its applications in finance arose in the early millennia, with applications to credit instruments and risk time-series soon after\cite{Prokhorenkova2018}.

When applied to financial volatility data, gradient boosting models, including LightGBM, could encounter significant challenges. Volatility time series are often characterised by events of rapid positive skew and potential out-of-distribution values that boosting relies on. this inability to extrapolate beyond the range can lead to poor generalisation\cite{Friedman2001}.

This limitation arises because gradient boosting models aggregate many decision trees and are fundamentally sets of piecewise functions. A gradient boosting model \( f(x) \) can be expressed as a sum of \( M \) individual decision trees \( T_m(x) \):

\[
f(x) = \sum_{m=1}^{M} \lambda_m T_m(x)
\]

where \( \lambda_m \) are the learning rates. Each decision tree \( T_m(x) \) partitions the space into regions \( R_{m,j} \) and assigns a constant value \( c_{m,j} \) to each region:

\[
T_m(x) = \sum_{j=1}^{J_m} c_{m,j} \mathbb{I}(x \in R_{m,j})
\]

Here, \( \mathbb{I}(x \in R_{m,j}) \) is an indicator function that equals 1 if \( x \) falls within region \( R_{m,j} \), and 0 otherwise.

As a result of the above, under-performance can be expected in high vol regimes. This could be alleviated with careful pre-processing and feature engineering, which we engage in to some extent.

We next consider SVMs, a quasi-linear classification model that has demonstrated efficacy in predicting vol, as highlighted in the literature. SVMs are frequently used in GARCH ensembles and in models where levels and returns are included as features\cite{Bezerra2017, Lin2021}, akin to ARCH. They have also been occasionally used to predict IV\cite{Gunnarsson2024}. One of the key reasons SVMs are favored are their multiple kernels, which can map relationships in arbitrarily high-dimensional spaces\cite{Scholkopf2002}. This approach is computationally efficient, as it allows SVMs to uncover complex nonlinearities in a low-dimensional and cost-effective manner.

SVMs have somewhat secured their place as a standard predictive model and are featured in the popular O'Reilly textbook Machine Learning for Financial Risk Management by Abdullah Karasan\cite{Karasan2022}
. Additionally, SVMs have been applied in various other significant areas, often as classifiers, including the prediction of default risk\cite{Chen2018}, demonstrating their empirical strength in a range of tasks. We have included SVMs as a benchmark of sorts for this reason.

We finally turn our \textit{attention} to ANNs broadly. The architectures of ANNs vary significantly, and we explore the idiosyncrasies of our model in the methodology section. ANNs have been applied to financial time-series as early as the 1980s, during one of the early waves of AI \cite{White1988}. It is argued that constraints surrounding hardware, data quality, and the lack of forms of regularisation, optimisation, suitable activation functions, or methods to address exploding gradients limited this early progress \cite{Saad1998}. Recently, however, renewed interest in ANNs has led to a wave of new research in their applications to financial time-series, likely driven by the alleviation of such constraints together with new architectures, such as RNNs. This has also been fuelled by applications in market microstructure, forecasting by hobbyists, and increased academic interest \cite{Heaton2017}.

In the context of vol prediction, various ANN architectures have been explored, ranging from comprehensive hybrid Recurrent Neural Networks (RNNs) to more basic Multi-Layer Perceptrons (MLPs). For instance, Gewenbo et al. \cite{Gewenbo2023} found strong performance in Historical Volatility (HV) and Realised Volatility (RV) forecasting using CNN-LSTMs. However, for Implied Volatility (IV), the naïve model outperformed others, achieving the smallest error across all five assets analyzed in their study. Despite this, the CNN-LSTM and MLP models showed promising results, although additional features did not significantly enhance predictive ability.

Liu \cite{Lin2021} further demonstrated the effectiveness of deep learning models like LSTMs for volatility prediction, showing that LSTM models outperformed classical methods such as GARCH and Support Vector Machines (SVMs). However, traditional deep learning models like LSTMs and GRUs\cite{Cho2014} are limited by their memory capacity, where valuable remote information can degrade as it is transmitted acrosss steps, potentially leading to prediction errors.

In addition, hybrid models combining GARCH with ANNs date back to 1997\cite{Donaldson1997}
 and approaches combining GRUs have been explored for RV forecasting across multiple asset classes. Michańków et al. \cite{Michankow2020} conducted extensive experiments using these hybrid models and reported improvements in prediction accuracy by leveraging the strengths of both GARCH models' theoretical rigor and the DL flexibility of GRUs.

Attention mechanisms have sparked interest in time-series forecasting literature. Lim et al. \cite{Lim2021} introduced the TFT, which combines attention mechanisms and RNN structures in its architecture. A recent comprehensive study from Nixtla\cite{Garza2023}, as well as work by Lim et al. \cite{Lim2021}, demonstrated that TFT achieves consistent performance across single-step and multi-step forecasts in a range of sequence prediction tasks.

Recent findings\cite{Frank2023} have indicated that TFTs are effective in predicting Realised Volatility (RV) and can outperform LSTMs and Random Forests when using pooling methods, with the results again being robust across different training methods. Similarly, Dai et al. \cite{Dai2019} and Wen et al. \cite{Wen2023} have highlighted that numerous transformer variations have shown promising results \cite{Farsani2021}.

The `attention' mechanism's proficiency in discerning the importance of time-dependent relationships within data makes it particularly relevant for market volatility forecasting \cite{Frank2023,Hu2021,Olorunnimbe2022}, where the ability to adapt to meaningful changes in states from a a feature matrix \cite{Lim2021}, or `highlight' endogenous volatility clustering could aid this success. The TFT model’s recent adoption in financial applications and its reported success in various empirical studies for volatility prediction \cite{Frank2023,Hu2021,Lim2021,Olorunnimbe2024} further highlight its potential as a powerful tool for this research.

\section{Methodology}
\label{sec:methodology}
\subsection{Data Description}
The raw data has a daily resolution and is derived from several sources. Our target vector \(\mathbf{y}\), as discussed extensively, is the level of the iTraxx/Cboe Europe Main 1-Month Volatility Index (BP Volatility). The feature matrix \(\mathbf{X}\) comprises price and volume data from Euro-denominated accumulation ETFs as well as short-term interest rate futures (Euribor c2 and €STR). together with other features (comprehensively outlined in the appendix) which did not make it to the primary set for further experimentation.

Raw features were meticulously selected by leveraging domain knowledge and represent a sample of components that act as proxies for inputs commonly seen in closed-form solutions like the Merton\cite{Merton1974} risk model (levels of debt, equity, risk, rates). The broad set of features undergo further processing and transformations, including a final dimensionality reduction step.

\subsection{Candidate Models}

Models were selected based on our literature review together with prior experimentation. The three chosen models consequentially ended up representing distinct supervised learning approaches. The Support Vector Machine (SVM) is a classical model that operates as a low-parameter classifier in a high-dimensional space. Gradient boosting, on the other hand, is a high-parameter ensemble method typically applied to lower-dimensional feature spaces, and lastly, the proposed ANN, representing a contemporary approach, includes three key components: Bidirectional gated recurrent units (GRUs) with multi-headed attention and residual connections. We further detail model specifics below.

We will refrain from further explanations with respect to model types as this was thoroughly examined in the literature review, but we will outline the model particulars and quirks of the architectures.

Outside of adaptation from typical use with categorical features achieved by the sequencing of data, and the incremental batch learning approach/hyperparameter selection as extensively outlined in the following sections, both the SVM and LightGBM models do not undergo significant changes from their standard architectures. It is perhaps worthy of note that the LightGBM, in contrast to conventional gradient boosting, grows trees leaf-wise. This means it splits the leaf with the highest loss reduction at each iteration\cite{Ke2017}. This potentially allows it to capture intricate patterns in non-iid time-series data, albeit with higher complexity.

\subsection{ATTN-GRU Architecture}

The ANN is slightly more sophisticated, and we will leave extensive discussion of layers and mechanisms outside the scope of this paper and instead cite seminal or relevant work and discuss the composition, which is summarised in Fig. 1.

The input is a 1D CNN layer, CNN layers have been contentious with respect to nontraditional use, yet we found them more effective than both Temporal Convolutional Network (TCN) or Causal Convolutional layers. Notably, a recent gold-winning model\cite{nimashahbazi} in Optiver's 'Trading at the Close' Kaggle competition\cite{kaggle_optiver} used a CNN input layer to extract local features from time-series data. In our architecture, the large number (64) of channels can be seen as various permutations of the raw input features, which the subsequent layers then weigh. despite greater complexity than some time-series ANNs, the model maintains a fairly shallow structure, a favoured architecture\cite{Gewenbo2022} for vol which aligned with our experimentation.

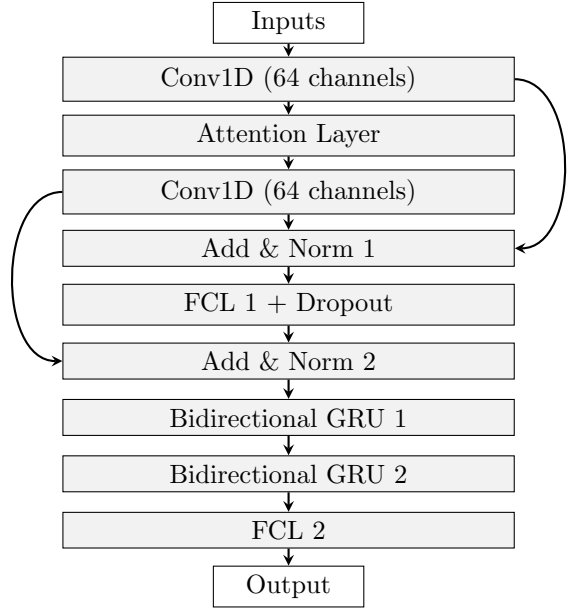
\begin{figure}[h]
    \centering
    \begin{tikzpicture}[node distance=0.75cm]

    \tikzstyle{startstop} = [rectangle, minimum width=2cm, minimum height=0.25cm, text centered, draw=black, fill=white]
    \tikzstyle{process} = [rectangle, minimum width=6cm, minimum height=0.25cm, text centered, draw=black, fill=gray!10]
    \tikzstyle{arrow} = [thick,->,>=stealth]

    \node (start) [startstop] {Inputs};
    \node (conv1) [process, below of=start] {Conv1D (64 channels)};
    \node (attn) [process, below of=conv1] {Attention Layer};
    \node (conv2) [process, below of=attn] {Conv1D (64 channels)};
    \node (addnorm1) [process, below of=conv2] {Add \& Norm 1};
    \node (fc1) [process, below of=addnorm1] {FCL 1 + Dropout};
    \node (addnorm2) [process, below of=fc1] {Add \& Norm 2};
    \node (gru1) [process, below of=addnorm2] {Bidirectional GRU 1};
    \node (gru2) [process, below of=gru1] {Bidirectional GRU 2};
    \node (fc2) [process, below of=gru2] {FCL 2};
    \node (output) [startstop, below of=fc2] {Output};

    \draw [arrow] (start) -- (conv1);
    \draw [arrow] (conv1) -- (attn);
    \draw [arrow] (attn) -- (conv2);
    \draw [arrow] (conv2) -- (addnorm1);
    \draw [arrow] (addnorm1) -- (fc1);
    \draw [arrow] (fc1) -- (addnorm2);
    \draw [arrow] (addnorm2) -- (gru1);
    \draw [arrow] (gru1) -- (gru2);
    \draw [arrow] (gru2) -- (fc2);
    \draw [arrow] (fc2) -- (output);

    % residual connections
    \draw [arrow] (conv1.east) to[out=0,in=0] (addnorm1.east);
    \draw [arrow] (conv2.west) to[out=180,in=180] (addnorm2.west);

    \end{tikzpicture}
    \caption{A high-level overview of the ATTN-GRU data flow, including residual connections.}
    \label{fig:model_flowchart}
\end{figure}

following the local features extracted by the first Conv1D, an attention layer evaluates similarities across the sequence and weights pertinent observations more heavily. The second Conv1D layer refines these features by reintroducing a local focus after this global processing. The key idea here is to allow the model to consider both local and global information at different stages, ensuring that neither is overlooked, building flexibility into the model, which, in conjunction with the residual connections, allows a 'switch' if deployed for online learning.

The operation of this part of the architecture can be represented as follows:
\[
\mathbf{H}_1 = \text{CNN}(\mathbf{X}), \quad
\mathbf{A} = \text{ATTN}(\mathbf{H}_1), \quad
\mathbf{H}_2 = \text{CNN}(\mathbf{A})
\]
\[
\mathbf{Z}_1 = \mathbf{H}_1 + \mathbf{H}_2, \quad
\mathbf{O} = \text{LayerNorm}(\mathbf{Z}_1) + \text{FCL}(\mathbf{Z}_1)
\]

These residual connections also allow the model to retain and refine the original local features extracted by the first Conv1D, even after they have been globally processed by the attention mechanism and then further refined by the second Conv1D, making it an architectural ensemble of sorts.

We started off with something resembling a TFT (albeit without static features) but found that removing some components and adding a couple of GRU layers was comparable and sufficient. These GRUs further process the output from the attention layer. By using bidirectional\cite{schuster1997} GRUs, the model can keep and 'forget' dependencies/states in past time steps (in one cell) and future time steps (in another cell) during training.

The network architecture incorporates layer normalisation at various points to stabilise training, additionally, the architecture includes FCLs that progressively reduce the dimensionality of the data, resulting in a single scalar output which represents the last time step.

The heart of accuracy usually comes at optimisation, but we did not find any benefit in excessive tinkering. We settled with Adaptive Moment Estimation (Adam) and employ gradient clipping to prevent exploding gradients during training, and dropout to further reduce overfitting.

\subsection{Feature Engineering}

Our feature engineering approach includes standard transformations commonly found in financial machine learning literature.

For prediction purposes, we apply a transformation to the target vector, \( \mathbf{y} \), (the level of the Europe Main 1-Month Volatility Index). Specifically, we compute the log differences, \( \tilde{\mathbf{y}}_t = \log(\mathbf{y}_t) - \log(\mathbf{y}_{t-1}) \), which can be interpreted as the delta of each timestep in the series. This transformation helps manage the significant positive skew of the volatility series and mitigates nonstationarity in the data. A well-behaved error distribution is standard practice for many robust prediction tasks. This also assists with model requirements, such as those of gradient boosting mentioned earlier.

Similarly, we apply the log-differences transformation to the entire array of input features \(\mathbf{X}\), which approximates returns and has the same normalising effect. A given price or volume series is transformed as outlined above.

Next, we calculate the rolling 21-day realised volatility (RV) of differenced log prices. For a given window of returns \(\{r_{t-20}, r_{t-19}, \ldots, r_t\}\), the RV is defined as: \( \text{RV}_t = \left(\frac{1}{21} \sum_{i=t-20}^{t} (r_i - \bar{r})^2\right)^{1/2} \), where \( \bar{r} \) is the mean of the returns over the 21-day window.
 A 21-day period was chosen for multiple reasons, one, it is the standard trading month, also, in prior experimentation, such a window had the strongest relationship with the label.
\[
\mathbf{X}_{\text{engineered}} = \left[ P_t, r_t, \text{RV}_t \right]
\]

We then concatenate all series, creating a set of engineered features that includes the price levels, the log returns, and the RV. This approach ensures that our feature set \(\mathbf{X}_{\text{engineered}}\) includes a broad set of potential features representing the temporal dependencies and the volatility structure of the financial data, providing robust inputs for our models.

To further refine our feature set, we use feature importance scores from a Random Forest model, averaged over multiple time-series splits, to select the top 10 features. This was performed on data that significantly preceded any test sets (01/11/22 - 18/07/23) Actual security names have been replaced with more descriptive feature names. A comprehensive listing of experimentation periods/folds/sets together with detailed security names are in the appendix.
\[
\text{importance}_j = \frac{1}{n_{\text{splits}}} \sum_{i=1}^{n_{\text{splits}}} \text{importance}_{i,j}
\]

where \(\text{importance}_{i,j}\) is the importance of feature \(j\) in the \(i\)-th split.\\
\\
The Random Forest rank of engineered features is summarised in Table \ref{table:features}.
\begin{table}[h]
\caption{Random Forest Rank of Features}
\begin{tabular}{ll}
\hline
Feature Description \\
\hline
European Equities Volume (RV) * \\
Euro Short-Term Rate Volume ($\ln$ diffs) * \\
Global Mixed Debt Volume (RV) * \\
European Volatility Index VSTOXX (Levels) * \\
European Government Debt ($\ln$ diffs) * \\
European Corporate Debt ($\ln$ diffs) \\
Euribor Futures Volume (RV) \\
European Equities Volume (Levels) \\
European Equities ($\ln$ diffs) \\
Global Equities Volume (Levels) \\
\hline
\end{tabular}
\label{table:features}
\textit{Note: Features marked with an asterisk (*) denote inclusion in the primary feature set.}
\end{table}

\vspace{0.2cm}

It is notable that volume-related features and their transformations rank among the top attributes. This could suggest that market activity serves as a strong determinant of IV, potentially acting as a weak proxy for more granular order-book data. Recent research supports the importance of trading volume in volatility forecasting, particularly during periods of economic uncertainty \cite{Liu2022, Michael2024}, the selection of log-differences in Debt instruments is supported by the idea that volatility is dependent on the price return process\cite{Michael2024}. These 10 features constitute the final feature matrix used for experimentation in the spirit of incorporating a data-driven automated feature evaluation approach.

\subsection{Training and Validation}

Initially, a comprehensive parameter search was conducted on older data, and we progressively modified architectures during the 'Experimentation' phase (detailed in the appendix). Less complex models used a grid search with the aim of creating optimal parameter set for each model with respect to the MAE.
The Attention-GRU model is an exception due to the significant computational cost associated with neural architecture search (NAS) and hyperparameter searches for ANNs, its architecture, layers, units, batch size (within architecture), learning rate, and channels remain fixed and were arrived at via literature review and experimentation. It is worth noting that activation functions, parameters, and states in ANNs inherently introduce high variability upon retraining and such expressiveness can be thought of as rivalling the other two models.

\subsection{Parameter State Configurations}

The following represents the parameters available for model selection (for the non-ANN models) during training. This methodology enables dynamic parameter selection from a predefined set for each batch, facilitating optimal configurations for subsequent predictions which we outline further in this paper. This approach approximates a pseudo-online learning approach, with the model's internal states and strategies updating on a per-batch basis.

\vspace{0.6cm}
\noindent\textbf{SVM:} \\
\textit{Kernels:} poly, rbf, sigmoid \\
\textit{C:} 1.0 (default)
\textit{Gamma:} scale, auto, 0.1, 0.15, 0.2 \\
\textit{\(\epsilon\)-values:} 0.05, 0.1, 0.15) 
\textit{Loss:} \(\epsilon\)-insensitive, \\
\\
\textbf{LightGBM:} \\
\textit{Learning rate:} 0.005, \textit{Min gain to split:} 0.01 \\
\textit{Leaves:} 75, 100, 125, \textit{Min data in leaf:} 10, 20, 30 \\
\textit{Max depth:} -1, 5, 10, \textit{Feature fraction:} 0.4, 0.5, 0.6 \\
\textit{Bagging fraction:} 0.9, \textit{Freq:} 1 \textit{Loss:} MAE\\
\\
\textbf{ATTN-GRU:} \\
\textit{Learning Rate:} 0.07, \textit{Epochs:} 32 (per prediction) \\
\textit{Early Stopping:} (Patience of 5), \textit{Batch Size:} 32 \\
\textit{Conv1D filters:} (x2) 64 channel (dilation of 2) \\
\textit{Bidirectional GRUs:} 64 units (first layer, resulting in 128 outputs), 32 units (second layer, resulting in 64 outputs), \textit{Attention:} Head size: 16, heads: 4 \\
\textit{Loss:} MAE\\

The feature matrix, denoted as \(\mathbf{X} \in \mathbb{R}^{n \times m}\) where \(n\) is the number of observations and \(m\) is the number of features, is stacked with the target vector \(\mathbf{y} \in \mathbb{R}^{n}\) and then sequenced across 5 time-steps, as 5 approximates the trading week, assisting the model with seasonality, we initially felt that this was quite a small sequence length, but it proved to be robust during experimentation, indicating that latent regimes may be particularly short. Depending on the model architecture (classical ML or ANN), the input shape can be represented as either a flat matrix \(\mathbf{X}_{\text{flat}} \in \mathbb{R}^{k \times (m \cdot s)}\) for classical ML models, where \(k\) is the number of observations and \(s\) is the sequence length, or as a rank-3 tensor \(\mathbf{X}_{\text{tensor}} \in \mathbb{R}^{k \times m \times s}\) for ANN models.

Uniform noise is added during training across all models. This slight augmentation to each observation is drawn from a uniform distribution between -0.02 and 0.02.

We exercise fair treatment with the walk-forward (expanding window) validation method that splits the batch into training and testing sets iteratively, ensuring that the model is evaluated on unseen data at each step. Initially, the training set is constructed with \((\text{features} \times \text{sequences}) \times 10\) observations, denoted as \(\mathbf{X}_{\text{train}} \in \mathbb{R}^{10 \times m}\). The training set is incrementally increased after being validated by the next value \(n\) in the series, i.e., \(\mathbf{X}_{t+10+n}\), until the entire batch of data is exhausted. This process is repeated for every set of parameters. The most performant parameters are saved, and the model is then fit on the entirety of the data using these best parameters, which are subsequently used for prediction.

This maintains a somewhat standardised approach across different models, despite the ANN pipeline having inherent validation and batching. However, we aim to respect the idiosyncrasies of each model type with regard to dimensionality handling to ensure a fair assessment.

Our test set is the \( n \)-th time-step after the batch used for validation. Tests are conducted independently for each observation in the test set, resulting in each forecast being the outcome of a distinct batch learning process, including entirely different sets of parameters. Although this is computationally intensive, it effectively eliminates any leakage of prior training or parameter states. This meticulous process is also more true to a real-life scenario and removes incidental `luck' from specific local minima in the hyperparameter or error space, hopefully enhancing the robustness of the experimental results.
\section{Experiments and Results}
\label{sec:experiments}
\subsection{Setup}

The experiments were conducted on Google Colab Pro, using an L4 GPU instance with the following configuration:
\vspace{0.3cm}

\noindent\textbf{Hardware:} NVIDIA L4 GPU with 24 GB GPU RAM.

\noindent\textbf{Experiment Parameters:} We tested the three models over three training window sizes: 63, 126, and 252 observations, each followed by 63 rolling predictions, resulting in batches that approximate quarterly, semi-annual, and annual histories. A sequence length of 5 was used.

\noindent\textbf{Stack:} detailed in the appendix.

\subsection{Evaluation Metrics}
We assess performance on the stationary values using conventional metrics and separately assess performance on levels of volatility.
\vspace{0.3cm}

\noindent\textbf{Mean Absolute Error (MAE):} Assesses performance on values.

\noindent\textbf{Root Mean Squared Error (RMSE):} Assesses performance on values.

\noindent\textbf{Mean Absolute Percentage Error (MAPE):} Assesses the accuracy of the predictions as a percentage, making it suitable for scaled data.

\noindent\textbf{Log Loss (LL):} The logarithmic Loss function \cite{Pagan1990} penalizes volatility forecasts symmetrically during low and high volatility periods.
\vspace{0.6cm}

We evaluated model performance over 63 observations (14/05/24 - 09/08/24) and performed statistical significance calculations of the predictions on the raw target values using the Diebold-Mariano (DM) test\cite{diebold1995}. The DM test was conducted to compare the predictive accuracy against a Naïve estimate. Tests were conducted using the forecast horizon of ($63 = 1$) and a significance level of 5\%. A t-distribution was used as the sample distribution. Fig. 2 - 4 are plots presenting the dispersion of the errors for models trained over a 242-period (approximately a year in trading days) window.

\subsection{Results}
\begin{table}[h]
\centering
\caption{Performance and Significance}
\begin{tabular}{lccc}
\hline
Model & MAE & RMSE & DM Test \\
\hline
Naïve & 0.105 & 0.162 & - \\
ATTN-GRU & & & \\
\quad 63-period & 0.080 & 0.119 & 0.46 (0.65) \\
\quad 126-period & 0.078 & 0.119 & 0.61 (0.55) \\
\quad 252-period & 0.075 & 0.117 & 0.55 (0.59) \\
SVM & & & \\
\quad 63-period & 0.082 & 0.127 & 0.44 (0.66) \\
\quad 126-period & 0.092 & 0.148 & 0.13 (0.90) \\
\quad 252-period & 0.086 & 0.130 & -0.00 (1.00) \\
LightGBM & & & \\
\quad 63-period & 0.088 & 0.127 & 0.56 (0.58) \\
\quad 126-period & 0.085 & 0.125 & 0.26 (0.80) \\
\quad 252-period & 0.086 & 0.128 & 0.50 (0.62) \\
\hline
\end{tabular}
\label{table:performance_significance}
\end{table}

Perhaps unsurprisingly, the ATTN-GRU demonstrates the best performance over the larger window, this is consistent with expectations and aligns with Lim's findings\cite{Lim2021} that annual windows yield the best performance for TFT architectures. The LightGBM also demonstrates uniform performance across all windows, but pales in terms of accuracy vs. SVM.

\begin{table}[h]
\caption{Performance Metrics, Index Levels}
\begin{tabular}{llrr}
\hline
Model & Window & MAPE & Log Loss \\
\hline
Naïve & 1-period & 11.136 & 2.617 \\
ATTN-GRU & 63-period & 7.756 & 1.409 \\
" & 126-period & 7.565 & 1.412 \\
" & 252-period & \textbf{7.290*} & \textbf{1.362*} \\
SVM & 63-period & \textbf{7.912*} & \textbf{1.603*} \\
" & 126-period & 9.657 & 2.199 \\
" & 252-period & 8.578 & 1.695 \\
LightGBM & 63-period & 8.832 & 1.609 \\
" & 126-period & \textbf{8.596*} & \textbf{1.554*} \\
" & 252-period & 8.631 & 1.630 \\
\hline
\end{tabular}
\label{table}
\textit{Note: The Log Loss values have been scaled by $10^2$ for readability. The observations in the test set have a range of 57.00 - 21.14 (35.86) and a coefficient of variation of 28.4\%.}
\end{table}

\begin{figure}[ht]
    \centering
    \includegraphics[width=0.47\textwidth]{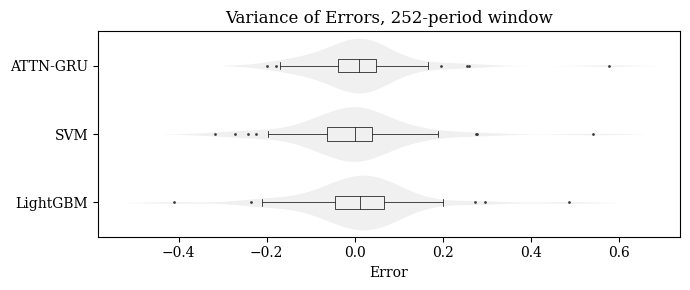}
    \caption{Variance of Errors across ATTN-GRU, SVM and LightGBM, respectively.}
    \label{fig:variance_of_errors}
\end{figure}

\begin{figure*}[t]
    \centering
    \includegraphics[width=1\textwidth]{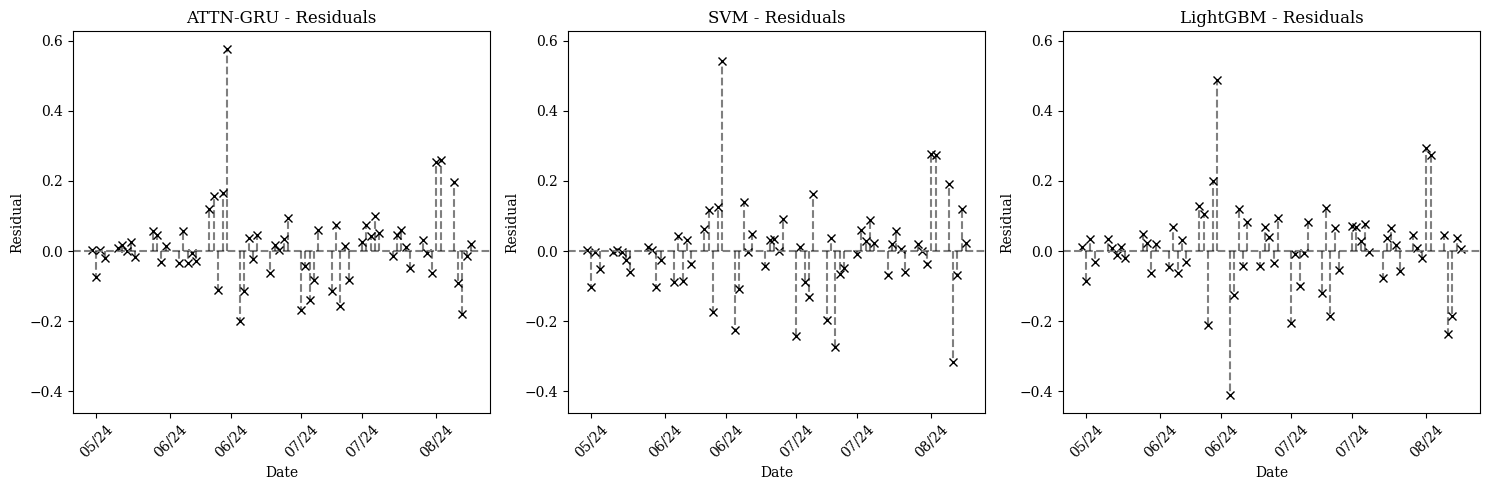}
    \caption{Residuals of the forecasts (Unscaled - units are in log differences).}
    \label{fig:residuals}
\end{figure*}

\subsection{Residual Analysis}
Upon cursory examination we can definitely see large errors occurring at extreme values, indicating that the models have been somewhat caught out during this dramatic underlying regime shift. The errors of LightGBM were potentially the most uniform across \( t \) and exhibited the least skew in its errors (visually) throughout the trials (This can be seen in Fig. 3). The sequential volatility spike occurring at the start of August, was handled particularly well across all models, which may be due to the incorporation of the similarly behaved patterns in the sequenced component as well as the expanding window within the batch.

\begin{figure*}[t]
    \centering
    \includegraphics[width=1\textwidth]{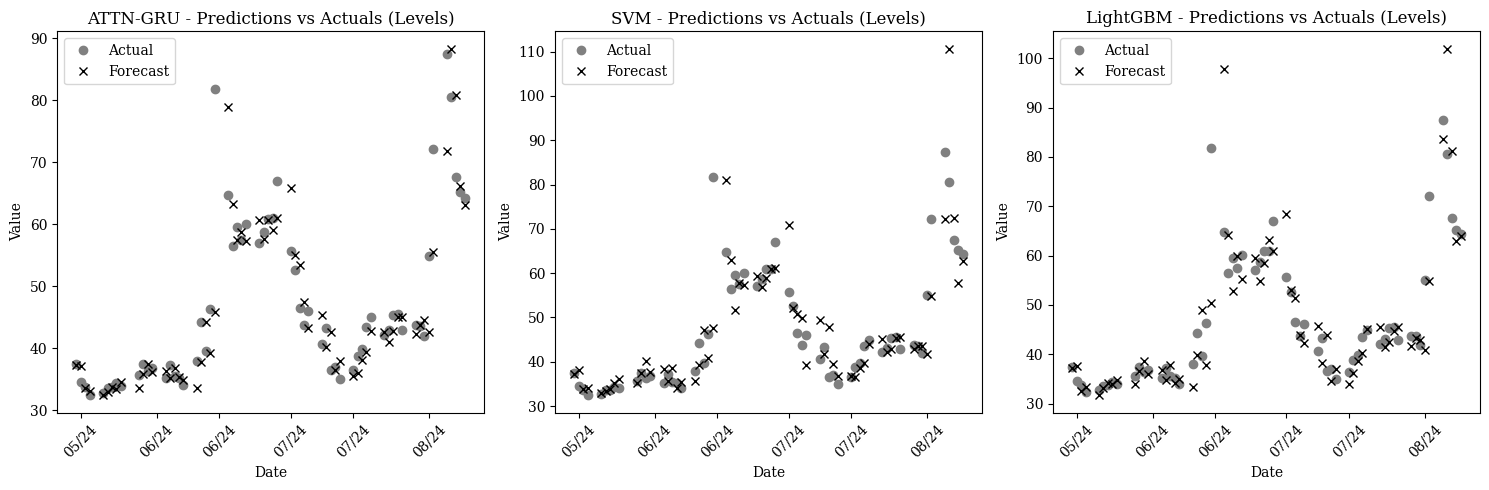}
    \caption{Predictions vs. Actuals (Scaled back to Index levels).}
    \label{fig:predictions_vs_actuals}
\end{figure*}

\section{Conclusion and Future Work}
\label{sec:conclusion}

\subsection{Summary}
Our research explored the effectiveness of three machine learning models—SVM, Gradient Boosting (LightGBM), and an Attention-GRU Hybrid in predicting the IV of the iTraxx/Cboe Europe Main 1-Month Volatility Index. The experimental results demonstrated that the Attention-GRU model slightly outperformed the other models across all error metrics, with all models exhibiting competitive performance versus naïve estimates, particularly in low-volatility periods, Ultimately, the models were not performant enough to exceed the rigour expected in rejecting the $H_0$ of our statistical tests. However, given that these tests were conducted on out-of-sample data, this does not entirely preclude the possibility of lower errors versus a naïve estimate going forward.

Key findings include the importance of feature engineering, particularly in transforming and normalising features and target. Additionally, the use of an Attention mechanism within the GRU architecture highlighted the potential of hybrid models in capturing complex time-series patterns, especially in the context of financial data where volatility clustering and mean-reverting behaviour are prevalent, however, associated parameter search and an appropriate optimisation set-up is crucial but also costly, so justifications for their use in on-the-fly applications are limited.

\subsection{Implications}
The findings of this research have several implications for financial risk prediction. The performance of the Attention-GRU hybrid model suggests that more expressive yet well-parameterised empirical models can offer advantages in forecasting tasks involving complex financial data. We noticed an increase in accuracy with a greater number of observations, as touched upon earlier, which demonstrates suitability for some tasks but few-shot or sparse data applications may be preferable if there are data volume or resource constraints. However, pre-training/fine-tuning on similar datasets could be employed.

The classical ML models, however, demonstrated robustness, indicating adaptability to underlying recent states and their efficacy in scenarios with limited data, which would be well suited for illiquid and OTC risk transfers and single-name contracts (like those which comprise the broader index), or as vol surface inputs and so on.

Lastly, the incremental batch learning approach used in this study, somewhat simulating an online learning scenario, offers a practical framework for deploying machine learning models in real-time trading or risk management systems. It allows for continuous model adaptation in response to new data; essential in dynamic financial markets, we believe this framework to be more robust and applicable than the prevalent conventions of 'backtesting' models on historical data.

\subsection{Future Work}

A preprocessing step that includes screening for features just before, or during training, may have proven to result in lower errors. Also, dynamically dropping features, in the style of a TFT, could have been applied to all models in multiple ways. Linear Discriminant Analysis (LDA) and Independent Component Analysis (ICA) were also explored and showed promise in improving model accuracy. and perhaps somewhat ironically, dimensionality increases, as presented in the TimesNet paper,\cite{Wu2023} could have enriched the feature set. However, some of these steps were not implemented in the final analysis due to the complexity of ensuring fair comparison across different model training sets.

Unscented Kalman Filters were also explored, and proved ineffective, however, this transformation into states likely may just require extensive changes to model parameters. The efficacy of such transformations cannot be explicitly ruled out. Further enhancements in feature engineering such as those shown by R. Ho and K. Hung, who applied empirical mode decomposition, (a technique to break a single univariate series into multiple terms, typically trend, seasonality, and noise) from time-series data before feeding it into a TFT architecture with gated residual blocks. Their research demonstrates a notable improvement of approx. 30\% compared to using unchanged input features \cite{HoHung2024}. Such pre-processing is common in econometrics, where methods like Seasonal Trend Loess (STL) decomposition is used extensively. Additionally, further hyperparameter tuning, especially with respect to the learning rate and optimisation (AdamW is promising) more broadly could of course enhance model performance. 

Future work on the theoretical and domain side could also explore the inclusion of explicit interaction terms between covariates, i.e. approximating the Merton model, or leveraging more insights from SV models to capture more complex relationships in the data. 

\section*{Acknowledgments}

The author extends gratitude to Dr. Yongxin Yang for his valuable guidance throughout this research. Appreciation is also given to the School of Electronic Engineering and Computer Science at Queen Mary, University of London, for providing essential resources and support. The author also thanks colleagues, friends, and family for their encouragement.\vspace{0.3cm}

\noindent This research was supported financially by the States of Guernsey.

\appendix

\subsection{Code and Data Availability}
The data, code, and outputs included in this paper together with extensive supplementary material is openly available on GitHub for the sake of transparency, replication, and further research. The repo can be accessed via the following link: [https://github.com/robtaylor94/Credit-VIX-CDS-IV-Prediction].

\subsection{Tools and Data Sources}

\textbf{NumPy and Pandas} \
C. R. Harris et al., Array programming with NumPy. Nature 585, 357–362 (2020) / The pandas development team \& McKinney W, others. Data structures for statistical computing in python. In: Proceedings of the 9th Python in Science Conference. 2010. p. 51–6.
\textit{Usage:} Data manipulation and preprocessing \
\textit{Application:} Extensively used throughout model development.

\noindent \textbf{SciPy and Statsmodels} \
Virtanen et al., SciPy 1.0: Fundamental Algorithms for Scientific Computing in Python. Nature Methods, 17(3), 261-272.
\textit{Usage:} Statistical Tests \
\textit{Application:} Applied for ACF for the Diebold-Mariano test and t-distributions for same \& GW test.

\noindent \textbf{Scikit-learn} \
Pedregosa et al., JMLR 12, pp. 2825-2830, 2011
\textit{Usage:} Machine Learning implementation \
\textit{Application:} Employed for SVMs, scalars, metrics, etc.

\noindent \textbf{LightGBM} \
©Copyright 2023, Microsoft Corporation
\textit{Usage:} Gradient boosting \
\textit{Application:} Used to improve model performance and prediction accuracy.

\noindent \textbf{PyTorch} \
Paszke et al., NIPS-W, 2017
\textit{Usage:} Neural network building and training \
\textit{Application:} Used specifically for constructing and training ATTN-GRU.

\noindent \textbf{Matplotlib and Seaborn} \
J. D. Hunter, "Matplotlib: A 2D Graphics Environment", Computing in Science \& Engineering, vol. 9, no. 3, pp. 90-95, 2007.
\textit{Usage:} Data visualisation \
\textit{Application:} Used for creating plots

\noindent \textbf{S\&P Global} \
\textit{Description:} iTraxx/Cboe Europe Main 1-Month Volatility Index (BP Volatility) \
\textit{Details:} Target data obtained from S\&P Global. This source has already been referenced in the main text.

\noindent \textbf{Stooq} \
\textit{Description:} Euro Currency Index data \
\textit{Details:} Sourced from Stooq, a comprehensive financial data provider.

\noindent \textbf{Investing.com} \\
\textit{Description:} Various Euro-denominated Accumulation UCITS ETFs and Futures data \\
\textit{Details:} Sourced from Investing.com, ISINs of which are shown below: \\

\noindent IE00BYX2JD69, IE00BDBRDM35, IE00B1YZSC51, IE00B3F81R35, IE00B4K48X80, LU0321462870, IE00BMQ5Y557, DE000A0C3QF1, EU000A2X2A254, EU0009652783

\subsection{Data Partitions}
\textbf{Initial Feature Selection} \
\textit{Time Period:} 01/11/22 - 18/07/23 \
\textit{Description:} Fit transformed features with Random Forest against lagged target.

\noindent \textbf{Experimentation} \
\textit{Time Period:} 01/11/22 - 14/07/23 \
\textit{Description:} Involved architecture search, parameter tuning, and cross-validation to refine model performance.

\noindent \textbf{Training Set} \
\textit{Time Period:} 16/05/23 - 14/05/24 \
\textit{Description:} Final training using final architectures and optimised hyparameters, this range covers the 63, 126 and 252 period windows.

\noindent \textbf{Test Set} \
\textit{Time Period:} 14/05/24 - 09/08/24 \
\textit{Description:} The model was evaluated on an out-of-sample test set to assess its generalisation performance and accuracy.

\end{document}